\documentclass[conference,a4paper]{IEEEtran}
\usepackage{graphicx}
\usepackage{amsmath}
\usepackage{amssymb}
\usepackage{booktabs}
\usepackage[hyphens]{url}
\usepackage[colorlinks=true,allcolors=blue]{hyperref}
\usepackage{algpseudocode}
\usepackage{algorithm}
\usepackage{tabularx}
\usepackage{multirow}

\usepackage{float}

\title{Optimal Base Station Placement for Beyond 5G Networks with Non-Convex Topology}
\author{
    \IEEEauthorblockN{ Mohamed Shalma, Amr Mansour,  and Ahmed El-Mahdy}
    \IEEEauthorblockA{Faculty of Information Engineering and Technology\\
    The German University in Cairo\\
    Cairo, Egypt\\
      mohamed.hamed@guc.edu.eg; amr.mansour@student.guc.edu.eg; ahmed.elmahdy@guc.edu.eg
    }
}

\begin{document}
\maketitle

\begin{abstract}
This paper investigates the optimal placement of a millimeter-wave (mmWave) base station (BS) within a realistic U-shaped environment with non-convex topology. The problem is challenging and NP-hard due to the non-convex topology and the non-convex objective functions which are the sum-rate maximization and max–min fairness, the latter being additionally non-smooth. To address this challenge, the BS placement is formulated as a Markov Decision Process (MDP). Then, we propose two deep reinforcement learning (DRL) techniques: First, the deployment area is discretized into a grid and optimized using a Deep Q-Network (DQN). Second, the U-shaped region is partitioned into continuous subspaces, where a Deep Deterministic Policy Gradient (DDPG) agent is dedicated to each subspace then the best BS placement is selected among partitions. Results demonstrate that optimal placement achieves full coverage and yields a Jain index of 0.99. Furthermore, the proposed partitioned multi-space DDPG achieves better solution than DQN with lower complexity.

\end{abstract}


\section{Introduction}
The transition toward Beyond 5G (B5G) and 6G wireless architectures is driven by the demand for ultra-high data rates, millisecond-level latency, and massive device connectivity. To meet these requirements, network operators are increasingly leveraging several technologies \cite{10525765, 10525785}, including millimeter-wave (mmWave) frequencies, which offer expansive bandwidths but present significant propagation challenges. Unlike sub-6 GHz signals, mmWave transmissions are highly susceptible to severe path loss and physical blockages from urban infrastructure. Consequently, the strategic placement of base stations (BSs) becomes a critical determinant of network performance, as even minor spatial deviations can lead to significant signal degradation. 

Traditional network planning often relies on static mathematical solvers or exhaustive search algorithms. However, in complex environments with non-convex constraints and high-dimensional search spaces, these methods become computationally prohibitive and often provide worse solutions compared to recent advancements in Artificial Intelligence (AI) \cite{11232082}. By modeling infrastructure deployment as a sequential decision-making problem, Deep Reinforcement Learning (DRL), in particular, enables an intelligent agent to navigate complex topological constraints and learn optimal spatial mappings through interaction with a simulated environment \cite{11313425}. 

This paper investigates the optimal placement of a mmWave base station within the C-Buildings complex at the German University in Cairo (GUC). By formulating the placement challenge as a Markov Decision Process (MDP), we utilize a Deep Q-Network (DQN) to optimize for two distinct objectives: Max-Min Fairness for equitable user coverage and Sum-Rate Maximization for aggregate system capacity. Our work contributes to the development of self-organizing B5G networks by demonstrating the efficacy of AI-driven spatial optimization in realistic, campus-scale environments.

The contributions of this paper are summarized as:
\begin{itemize}
\item We investigate the optimal placement of the BS in practical urban scenario using the mmWaves frequency for B5G networks. The practical permitted vicinity for BS placement is non-convex U-shaped topology leading to an NP hard non-convex optimization problem.
\item Beside the non-convex topology, We consider two non-convex objective functions to optimize: first, the sum-rate maximization, and second, the max-min rate optimization is not only non-convex but also non-smooth.  
\item To tackle this NP-hard problem, we propose two DRL-based approaches: first by transforming the non-convex topology to a grid and using a DQN agent. Second, by partitioning the U-shaped topology into continuous action spaces then using multiple DDPG agents for each partition. Both approaches lead to an optimal solution with considerable gain over standard vertex points.
\end{itemize}

\section{Literature Review}
Efficient 5G millimeter-wave (mmWave) base station (BS)
deployment is a major challenge. Unlike traditional cellular
systems, mmWave communication requires carefully opti-
mized BS placement to maintain reliable coverage and high
data rates. Recent studies highlighted that spatial geometry,
user distribution, and line-of-sight (LoS) conditions strongly
influence signal propagation and overall network performance
\cite{10.1177/1550147720926374,9045670,10942890}.

% Several studies investigated propagation behavior in com-
% plex wireless environments to improve deployment efficiency.
% These works showed that walls, metallic objects, and sur-
% rounding structures significantly affect signal strength and
% path loss. In addition, reflections from nearby surfaces can
% partially improve NLoS communication and enhance coverage
% performance in obstructed regions \cite{11232084}.

To address these challenges, machine learning and deep re-
inforcement learning (DRL) techniques have recently been in-
tegrated into wireless optimization frameworks. DRL enables
intelligent agents to dynamically learn optimal deployment
strategies through interaction with the environment without
relying on fixed mathematical models. Deep reinforcement
learning methods such as Deep Deterministic Policy Gradient
(DDPG) and other policy-based approaches have been used
for BS placement and resource optimization, showing strong
performance in adaptive wireless systems \cite{333,444}.

Multi-armed bandit (MAB) approaches also demonstrated
efficient adaptation to changing channel conditions and user
distributions while reducing computational complexity
\cite{10.1145/3573942.3574002}. In addition, classical and
evolutionary optimization methods such as particle swarm
optimization and genetic algorithms have also been applied
to BS placement problems, achieving competitive results in
structured environments \cite{111,555}.

Recent advances in multi-agent deep reinforcement learning
(MADRL) further improved wireless optimization compared to traditional
optimization methods in dynamic communication environ-
ments. Multi-objective optimization approaches have also
been proposed to jointly optimize coverage, throughput, and
localization accuracy in 5G BS deployment systems
\cite{10.1109/10597044,10942890}.

In addition to conventional BS deployment, several re-
searchers investigated unmanned aerial vehicle (UAV) and
aerial BS placement using DRL and ray-tracing techniques.
These studies demonstrated that accurate spatial modeling and
LoS analysis are essential for maintaining stable communi-
cation links in complex environments. UAV-assisted systems
and aerial base stations have also been optimized using re-
inforcement learning, heuristic algorithms, and energy-aware
models to improve coverage and efficiency \cite{11232083,11232084,777,666}.

% Reconfigurable intelligent surface (RIS)-assisted communi-
% cation has also emerged as a promising solution for improving
% wireless coverage. RIS technology enhances signal quality by
% controlling electromagnetic wave reflections toward users in
% weak coverage areas. Recent studies combined RIS-assisted
% networks with deep reinforcement learning algorithms such
% as Soft Actor-Critic (SAC) and Deep Deterministic Policy
% Gradient (DDPG) to optimize resource allocation and improve
% system performance under dynamic channel conditions
% \cite{11313425,10942890}.

Despite the significant progress achieved in recent years,
many existing studies still rely on simplified propagation as-
sumptions or focus mainly on open-area scenarios. Limited
research addresses complex practical geometries where users
are distributed in partially enclosed structures such as U-
shaped environments. Therefore, intelligent DRL-based op-
timization frameworks are needed to determine optimal BS
placement and maximize network performance in these chal-
lenging environments.
% -------------------------------------------------------------------------------
% BS placement in wireless networks has been widely investigated in recent years, particularly for heterogeneous networks, UAV-assisted systems, and dense 5G/6G deployments. Most existing approaches aim to improve coverage, capacity, and energy efficiency under simplified environmental assumptions. However, they generally do not explicitly consider non-convex spatial constraints or mmWave-specific propagation effects, which are crucial in high-frequency systems where blockage and line-of-sight conditions dominate performance.

% In \cite{111}, a cooperative placement framework for 4G and 5G base stations is proposed using an improved particle swarm optimization (PSO) algorithm combined with DBSCAN clustering. The method exploits building density information to guide deployment and maximize coverage. Although effective in heterogeneous urban scenarios, it focuses mainly on coverage maximization and does not explicitly model mmWave propagation characteristics such as severe blockage and directional beamforming. In addition, it does not account for irregular non-convex geometries where spatial confinement strongly impacts link quality.

In \cite{222}, a robust BS placement method is proposed under uncertain traffic demand using a column-and-constraint generation approach. While the formulation captures demand uncertainty and reduces deployment cost, it assumes predefined candidate BS locations and does not consider complex environmental geometry. Moreover, it is not tailored to mmWave systems where LOS dependency and blockage effects are dominant.

Overall, existing literature lacks a unified framework that jointly considers mmWave propagation characteristics and non-convex spatial constraints such as U-shaped environments, which is addressed in this work.

\begin{figure}[t]
    \centering
    \includegraphics[width=\linewidth]{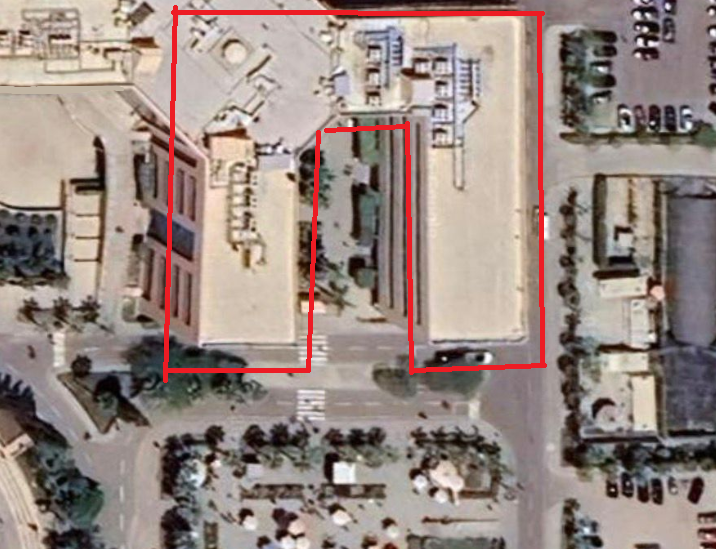}
    \caption{Satellite Image of the non-convex U-shaped topology of the C-building at the GUC campus }
    \label{fig:placeholder}
\end{figure}

\section{System and Channel Model}
\begin{figure}[!t]
    \centering
    \includegraphics[width=0.8\linewidth]{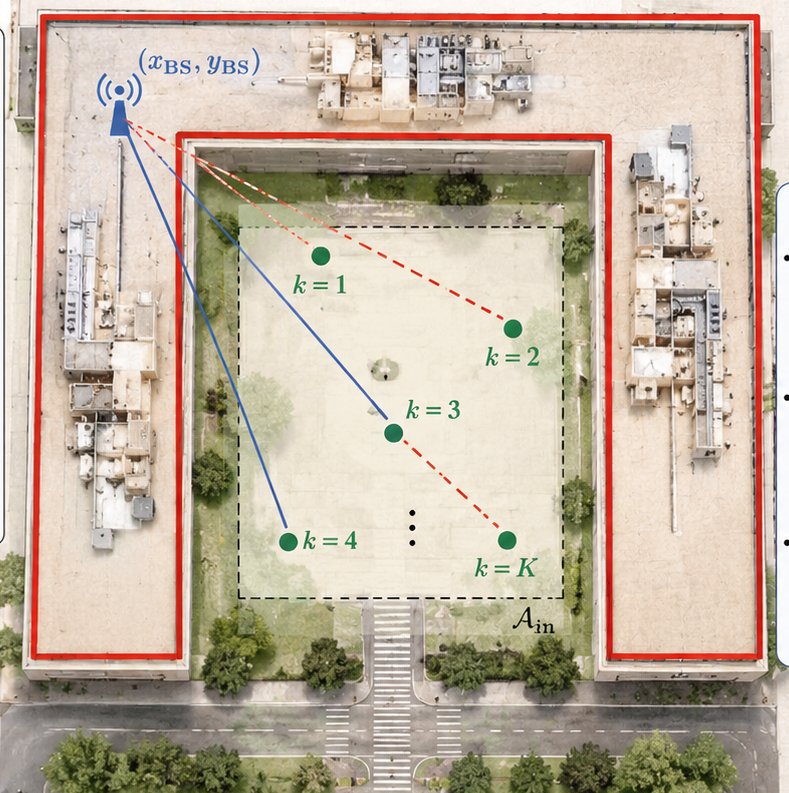}
    \caption{Vector Image of non-convex U-shaped topology of the C-building at the GUC campus}
    \label{fig:placeholder}
\end{figure}
\subsection{Model Description}
We consider a downlink mmWave communication scenario in which a single BS serves $K$ stationary user equipments (UEs) distributed within a defined indoor area. The environment is modeled after the C-Buildings complex at the GUC, where adjacent rectangular structures enclose an open courtyard. The BS is mounted on the exterior wall of a building at a fixed height, while UEs are located at ground level inside the courtyard. The objective is to determine the two-dimensional (2D) coordinates of the BS that optimize a chosen network performance metric, subject to physical deployment constraints. 

\subsection{Channel Model}
The Euclidean 3D distance between the base station and user $k$ is given by
\begin{equation}
d_k = \sqrt{(x_{\text{BS}} - x_k)^2 + (y_{\text{BS}} - y_k)^2 + (h_{\text{BS}} - h_k)^2}
\end{equation}
where $(x_{\text{BS}}, y_{\text{BS}})$ and $(x_k, y_k)$ denote the horizontal coordinates of the BS and user $k$, respectively, while $h_{\text{BS}}$ and $h_k$ represent their heights. The mmWave channel between the BS and user $k$ is modeled as a combination of large-scale path loss and small-scale fading:
\begin{equation}
h_k = \sqrt{L(d_k)} \, g_k
\end{equation}
where $g_k \sim \mathcal{CN}(0,1)$ represents Rayleigh fading. The large-scale path loss is modeled as
\begin{equation}
L(d_k) = C_0 d_k^{-\alpha}
\end{equation}
where $C_0$ is the path loss at a reference distance of 1 meter, and $\alpha$ is the path loss exponent. The received signal at user $k$ is expressed as
\begin{equation}
y_k = \sqrt{P_{\text{tx}}} \, h_k x + n_k
\end{equation}
where $P_{\text{tx}}$ is the transmit power, $x$ is the transmitted symbol with unit power, and $n_k \sim \mathcal{CN}(0, N_0)$ denotes additive white Gaussian noise. Assuming a noise-limited mmWave system, the signal-to-noise ratio (SNR) at user $k$ is given by
\begin{equation}
\gamma_k = \frac{P_{\text{tx}} |h_k|^2}{N_0}
\end{equation}
The achievable data rate for user $k$ is computed as
\begin{equation}
R_k = \log_2 \left(1 + \gamma_k \right)
\end{equation}

\section{Proposed DRL-Based BS Placement}
In this section, we provide the MDP formulation and the details of the proposed DQN and DDPG agents.
\subsection{MDP Modeling}
The DRL-based solution is mainly composed of observations, action, the DRL algorithm, and the reward function. Both agents share these MDP components in the same way which are further illustrated as:
\begin{itemize}
\item \textbf{Observations}: both agents take the users 2D locations as observations which are passed as  $\texttt{obsv}=[(x_1,y_1),(x_2,y_2),\cdots,(x_K,y_K)]$.
\item \textbf{Action Space}: both agents output the optimal location of the BS coordinates and the action vector is defined as $a_t=[x_{BS},y_{BS}]$.
\item \textbf{Reward}: The reward is given each time the agent selects an action that should reflect the objective function. We use the sum rate and minimum rate to exactly reflect the system objective. $\texttt{reward}_{\text{sum}}= R_{\text{sum}} = \sum_{k=1}^K \log_2(1+\gamma_k)$, $\texttt{reward}_{\text{fairness}}= R_{\text{fairness}} = \min_k \log_2(1+\gamma_k)$
\end{itemize}

\subsection{Vicinity Discretization using DQN}
The DQN works by discretizing the whole U-shape vicinity then observing the users' locations and then choosing an action as described in Algorithm 1 in a way that minimizes the loss function. The DQN agent approximates the Q-function using a fully connected neural network with $L$ layers and $N$ neurons per layer. To evaluate its complexity, the dominant computational cost arises from forward and backward propagation during training. The per-update complexity is given by
\begin{equation}
\mathcal{O}_{\text{DQN}} = \mathcal{O}(L N^2)
\end{equation}

Considering a replay buffer with mini-batch size $B$ and a total of $T$ training steps, the overall training complexity becomes
\begin{equation}
\mathcal{O}_{\text{DQN}}^{\text{total}} = \mathcal{O}(T \cdot B \cdot L N^2)
\end{equation}
Additionally, the action selection step requires evaluating all grid points. If the grid size is $|\mathcal{G}|$, the action selection complexity per step is
\begin{equation}
\mathcal{O}(|\mathcal{G}|)
\end{equation}
Thus, the total complexity of DQN can be expressed as
\begin{equation}
\mathcal{O}_{\text{DQN}} = \mathcal{O}(T \cdot B \cdot L N^2 + T \cdot |\mathcal{G}|)
\end{equation}
\begin{algorithm}[H]
\caption{BS Location Optimization using DQN (Discrete Grid)}
\begin{algorithmic}[1]
\State \textbf{Input:} User locations $\{(x_k,y_k)\}_{k=1}^K$, grid $\mathcal{G}$
\State Initialize Q-network $Q_\theta$ and target network $Q_{\bar{\theta}}$
\State Initialize replay buffer $\mathcal{D}$

\For{each episode}
    \State Initialize environment (user positions and channels)
    \For{each time step $t$}
        \State Observe state $s_t$
        \State Select action $a_t \in \mathcal{G}$ using $\epsilon$-greedy:
        \[
        a_t = 
        \begin{cases}
        \text{random grid point}, & \text{with prob. } \epsilon \\
        \arg\max_a Q_\theta(s_t,a), & \text{otherwise}
        \end{cases}
        \]
        \State Set BS location $(x_{\text{BS}},y_{\text{BS}}) \leftarrow a_t$
        \State Compute reward $r_t = R_{\text{sum}}$ or $R_{\text{fairness}}$
        \State Observe next state $s_{t+1}$
        \State Store $(s_t,a_t,r_t,s_{t+1})$ in $\mathcal{D}$
        \State Sample minibatch from $\mathcal{D}$
        \State Update $Q_\theta$ using Bellman loss:
        \[
        y = r_t + \gamma \max_{a'} Q_{\bar{\theta}}(s_{t+1},a')
        \]
        \State Periodically update target network:
        \[
        \bar{\theta} \leftarrow \theta
        \]
    \EndFor
\EndFor

\State \textbf{Output:} Optimal grid location $(x_{\text{BS}}^*,y_{\text{BS}}^*)$
\end{algorithmic}
\end{algorithm}
\subsection{Multi-Partitioning using DDPG}
In the multi-partitioned DDPG approach, the environment is divided into $3$ regions, where each region is assigned an independent DDPG agent consisting of an actor and a critic network as shown in Algorithm 2. Each DDPG agent requires updating two neural networks. Therefore, the per-update complexity per agent is
\begin{equation}
\mathcal{O}_{\text{DDPG, per-agent}} = \mathcal{O}(2 L N^2) \approx \mathcal{O}(L N^2)
\end{equation}
Considering $3$ partitions, the total per-update complexity becomes
\begin{equation}
\mathcal{O}_{\text{DDPG}} = \mathcal{O}(3 \cdot L N^2)
\end{equation}
Over $T$ training steps and mini-batch size $B$, the overall complexity is
\begin{equation}
\mathcal{O}_{\text{DDPG}}^{\text{total}} = \mathcal{O}(T \cdot B \cdot 3 \cdot L N^2)
\end{equation}
Unlike DQN, DDPG operates in a continuous action space and does not require exhaustive action search over a grid, which eliminates the $\mathcal{O}(|\mathcal{G}|)$ term.

\begin{algorithm}[H]
\caption{BS Location Optimization using Multi-DDPG (Partitioned Space)}
\begin{algorithmic}[1]
\State \textbf{Input:} User locations $\{(x_k,y_k)\}_{k=1}^K$, regions $\{\mathcal{A}_i\}_{i=1}^3$
\For{each region $i \in \{1,2,3\}$}
    \State Initialize actor $\pi_{\theta_i}$, critic $Q_{\phi_i}$
    \State Initialize target networks $\pi_{\bar{\theta}_i}, Q_{\bar{\phi}_i}$
    \State Initialize replay buffer $\mathcal{D}_i$
\EndFor

\For{each episode}
    \State Initialize environment
    \For{each region $i \in \{1,2,3\}$}
        \For{each time step $t$}
            \State Observe state $s_t$
            \State Select action with exploration:
            \[
            a_t^{(i)} = \pi_{\theta_i}(s_t) + \mathcal{N}_t
            \]
            \State Project action into region:
            \[
            a_t^{(i)} \leftarrow \text{Proj}_{\mathcal{A}_i}(a_t^{(i)})
            \]
            \State Set BS location $(x_{\text{BS}},y_{\text{BS}}) \leftarrow a_t^{(i)}$
            \State Compute reward $r_t^{(i)} = R_{\text{sum}}$ or $R_{\text{fairness}}$
            \State Observe next state $s_{t+1}$
            \State Store $(s_t,a_t^{(i)},r_t^{(i)},s_{t+1})$ in $\mathcal{D}_i$
            
            \State Sample minibatch from $\mathcal{D}_i$
            \State Update critic:
            \[
            y = r_t^{(i)} + \gamma Q_{\bar{\phi}_i}(s_{t+1}, \pi_{\bar{\theta}_i}(s_{t+1}))
            \]
            \State Update actor using policy gradient
            \State Soft update:
            \[
            \bar{\theta}_i \leftarrow \tau \theta_i + (1-\tau)\bar{\theta}_i
            \]
            \[
            \bar{\phi}_i \leftarrow \tau \phi_i + (1-\tau)\bar{\phi}_i
            \]
        \EndFor
    \EndFor
\EndFor

\State Select best region and location:
\[
(x_{\text{BS}}^*,y_{\text{BS}}^*) = \arg\max_{i,t} r_t^{(i)}
\]

\State \textbf{Output:} Optimal BS location $(x_{\text{BS}}^*,y_{\text{BS}}^*)$
\end{algorithmic}
\end{algorithm}
It is worth mentioning that the DQN approach suffers from scalability issues due to the discrete action space, as its complexity grows linearly with the grid size $|\mathcal{G}|$. In contrast, the multi-partitioned DDPG approach scales with the number of partitions $M$, making it more suitable for continuous optimization in large search spaces. 
\section{Numerical Results}
In this section, we provide numerical results based on software simulation of the proposed approaches. The simulation parameters are provided in TABLE I. The evaluation relies on key performance indicators, including minimum, mean, and maximum achievable rates (measured in Mbps), alongside Jain's fairness index and percentage coverage. These comparative results validate the efficacy of continuous action-space models in fine-tuning spatial parameters for complex wireless resource allocation scenarios.

\begin{table}[H]
\centering
\caption{ Simulation Parameters}
\label{tab:sim_params}
\begin{tabular}{|p{0.63\columnwidth}|p{0.25\columnwidth}|}
\hline
\multicolumn{1}{|c|}{\textbf{Parameter Description}} & \multicolumn{1}{c|}{\textbf{Value}} \\
\hline
\multicolumn{2}{|l|}{\textbf{Environment and Grid Constraints}} \\
\hline
Number of Users (\(K\)) & \(60\) \\
\hline
Cell Size & \(1\)\,m \\
\hline
Transmit Power (\(P_{tx}\)) & \(23\)\,dBm \\
\hline
Noise Power & \(-95\)\,dBm \\
\hline
Path Loss Exponent (\(\alpha\)) & \(3.5\) \\
\hline
Channel Bandwidth & \(20\)\,MHz \\
\hline
Carrier Frequency & \(10\)\,GHz \\
\hline
Maximum Steps per Episode & \(200\) \\
\hline
Training Timesteps & \(30{,}000\) \\
\hline
DDPG Learning Rate & \(1 \times 10^{-3}\) \\
\hline
Soft Update Factor (\(\tau\)) & \(0.005\) \\
\hline
DQN Learning Rate & \(5 \times 10^{-4}\) \\
\hline
Replay Buffer Size & \(100{,}000\) \\
\hline
Batch Size & \(128\) \\
\hline
Discount Factor (\(\gamma\)) & \(0.99\) \\
\hline
Network Architecture & MLP \([256,256]\) \\
\hline
Evaluation Episodes & \(10\) \\
\hline
\multicolumn{2}{|l|}{\textbf{User Spatial Distribution (Poisson Process)}} \\
\hline
Poisson Parameter (\(\lambda\)) & \(200\) \\
\hline
Mean User X-Position (\(\mu_x\)) & \(0\)\,m \\
\hline
Mean User Y-Position (\(\mu_y\)) & \(-15\)\,m \\
\hline
Spatial Standard Deviation (\(\sigma = \sqrt{\lambda}\)) & \(\approx 14.14\)\,m \\
\hline
\multicolumn{2}{|l|}{\textbf{Telecommunications and Channel Model}} \\
\hline
Transmit Power (\(P_{tx}\)) & \(1.0\)\,W (\(30\)\,dBm) \\
\hline
Noise Power (\(N_0\)) & \(1 \times 10^{-10}\)\,W \\
\hline
Path Loss Exponent (\(\alpha\), mmWave) & \(4\) \\
\hline
\end{tabular}
\end{table}

\begin{figure}
    \centering
    \includegraphics[width=\linewidth]{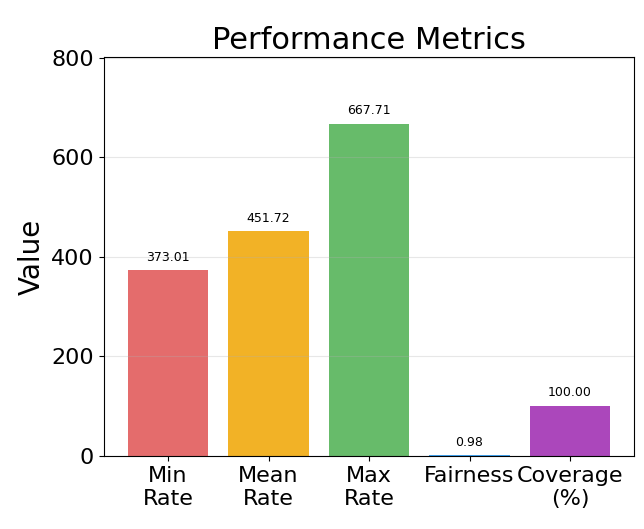}
    \caption{Performance metrics of DDPG}
    \label{pmddpg}
\end{figure}
\begin{figure}
    \centering
    \includegraphics[width=\linewidth]{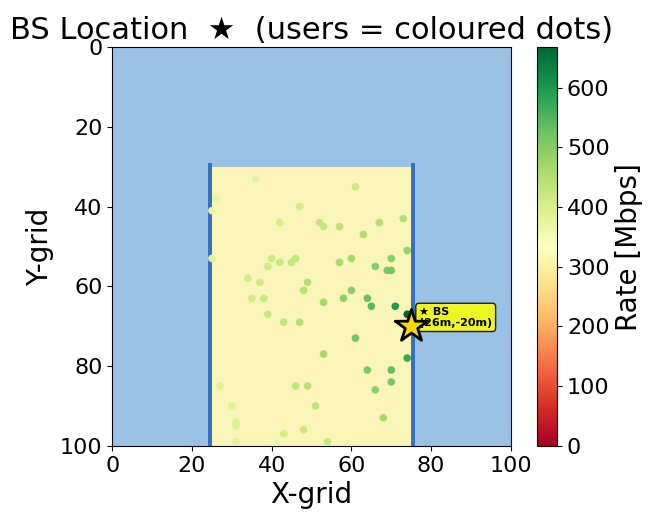}
    \caption{Optimal BS placement using DDPG}
    \label{blddpg}
\end{figure}

\begin{figure}
    \centering
    \includegraphics[width=\linewidth]{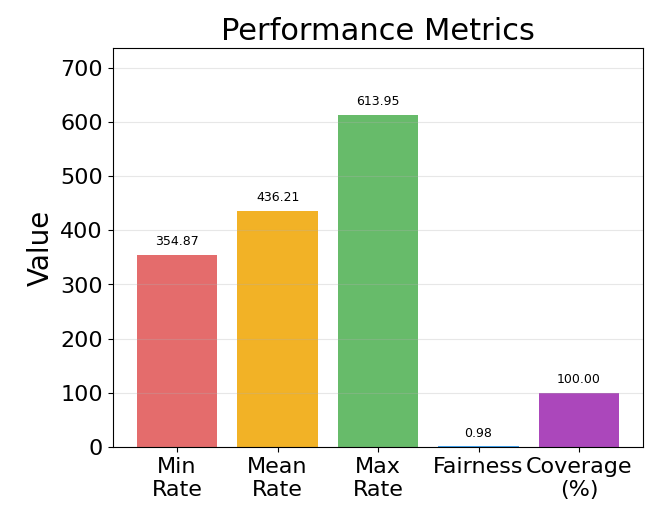}
    \caption{Performance metrics of DQN}
    \label{pmdqn}
\end{figure}

\begin{figure}
    \centering
    \includegraphics[width=\linewidth]{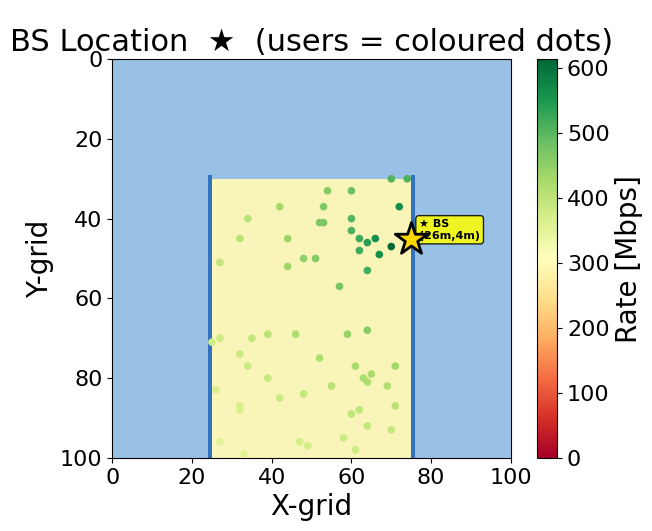}
    \caption{Optimal BS placement using DQN}
    \label{bldqn}
\end{figure}

The performance metrics for the DDPG algorithm is shown in Fig. \ref{pmddpg},  which demonstrate its strong capability in optimizing the base station location for maximum throughput. The bar chart reveals that the DDPG agent secures a minimum user rate of 373.01 Mbps, ensuring a robust baseline quality of service for edge users. Furthermore, it achieves an impressive mean data rate of 451.72 Mbps and peaks at a maximum rate of 667.71 Mbps. Notably, the algorithm accomplishes this high capacity while maintaining an excellent Jain's fairness index of 0.98 and guaranteeing 100\% network coverage for all simulated users in the environment.

The spatial distribution and the resulting user data rates governed by the DDPG agent are illustrated in Fig. \ref{blddpg}. As depicted in the figure, the continuous action-space formulation allows the DDPG algorithm to fine-tune the base station coordinates without being restricted by grid quantization, ultimately converging on a highly strategic location. Commentary on this visual layout highlights a highly optimal rate distribution; a significant cluster of users experiences premium data rates—represented by the dark green indicators exceeding 500 Mbps. This precise spatial positioning directly contributes to the superior mean and maximum throughputs observed in the preceding performance metrics.

The performance metrics for the discrete-action DQN algorithm are presented in Fig. \ref{pmdqn}. While the DQN agent successfully matches the DDPG approach by providing complete 100\% coverage and an identical, near-perfect fairness score of 0.98, its overall throughput is comparatively constrained. Specifically, the DQN agent yields a lower minimum rate of 354.87 Mbps, a mean rate of 436.21 Mbps, and a maximum rate of 613.95 Mbps. This relative drop in network performance underscores the inherent limitations of using discrete spatial actions when attempting to maximize capacity in a complex, multi-user topography.

The spatial layout resulting from the DQN agent's optimization is shown in Fig. \ref{bldqn}. While this centralized positioning along the Y-axis provides highly equitable coverage—as evidenced by the sustained fairness index—it yields a slightly less optimal rate distribution compared to the DDPG approach. 

\section{Conclusion}
In this paper, we investigated the optimal BS placement at mm-Waves frequency within a non-convex U-shaped topology aiming to maximize both sum-rate and maximum fairness. The proposed DRL approaches including DQN and multi-partitioned DDPG find a optimal solution while the DDPG finds a slightly better solution with even lower complexity. Finally, both approaches yield a Jain's index above 0.9 and $100 \%$ full coverage for all users.

\bibliographystyle{IEEEtran}
\bibliography{Ref}

\end{document}